\documentclass[12pt]{iopart}

\usepackage{iopams}  
\usepackage{multirow}
\usepackage{cleveref}
\usepackage{amssymb}
\usepackage{longtable}
\usepackage{booktabs}
\usepackage{tabu}
\usepackage{tabularx} 
\usepackage{graphicx}
\usepackage{inputenc}
\usepackage{url}
\usepackage{geometry}
\usepackage{fontenc}
\usepackage{xcolor}
\usepackage{hyperref}
\usepackage[colorinlistoftodos]{todonotes}
\usepackage{acronym}
\usepackage{subcaption}
\usepackage{array}
\usepackage{booktabs}
\usepackage[compress]{cite}
\usepackage{array}
\usepackage[normalem]{ulem}
\usepackage[switch]{lineno}
\usepackage{listings}

\usepackage{longtable}
\usepackage{tabularx}
\usepackage{tabulary}
\usepackage{booktabs}
\usepackage{multirow}
\usepackage{adjustbox}
\usepackage{breqn}
\usepackage{graphicx}
\usepackage{lipsum}% example text
\newcommand{\revised}[3]{#2}

\usepackage{xr}
\usepackage[switch]{lineno}
\begin{document}
\title{The Impact of On-chip Communication on Memory Technologies for Neuromorphic Systems}

\author{Saber Moradi and Rajit Manohar}

\address{Computer Systems Lab, School of Engineering~\&~Applied Science\\ Yale University, New Haven, CT 06520, USA}
\ead{first-name.last-name@yale.edu}
\vspace{10pt}
\begin{indented}
\item[]Sep 2018
\end{indented}

\begin{abstract}Emergent nanoscale non-volatile memory technologies with high integration density offer a promising solution to overcome the scalability limitations of CMOS-based neural networks architectures, by efficiently exhibiting the key principle of neural computation. Despite the potential improvements in computational costs, designing high-performance on-chip communication networks that support flexible, large-fanout connectivity remains as daunting task.
 In this paper, we elaborate on the communication requirements of large-scale neuromorphic designs, and point out the differences with the conventional network-on-chip architectures. We present existing approaches for on-chip neuromorphic routing networks, and discuss how new memory and integration technologies may help to alleviate the communication issues in constructing next-generation intelligent computing machines.

\end{abstract}

%\noindent{\it Keywords}: network-on-chip, communication, neuromorphic, emergent memory, memristor 
%\submitto{Journal of Physics D: Applied Physics}
%\maketitle
\section{Introduction}
Neuromorphic computing is an inter-disciplinary area of research,
aiming to emulate the computational principles of Biological neural
systems and to utilize such principles to solve complex and
computationally intensive problems. The discipline originated with the
goal of constructing silicon models of Biological neurons and
synapses, based on the observation that the basic Physics governing the
flow of current across a semiconductor junction and Biological
``junction'' in an ion channel were the same~\cite{Mead90}. Early papers in the field
include the design of silicon models of the retina and cochlea, with
analog CMOS circuits emulating Biological circuits discovered
through anatomical studies~\cite{Mahowald88,Mead_Tobi91,Tobi93,cochlea,park15}.

Concurrently, neural networks were being used for pattern recognition, where
neurons were modeled using the McCulloch-Pitts model~\cite{McCulloch}. In this
approach, a neuron is a function that partitions points in space into two
regions by a separating hyperplane. Mathematically, a neuron computes
a dot product between its weight vector and an input vector, and the
output is given by the sign of the result. Generalizations of this
where the output is computed using sigmoids or other functional forms
were also studied~\cite{activation}. More recently, such neural
networks with a large number of parameters (millions, if not more)
have been used to achieve unsurpassed classification accuracy for
tasks such as object labeling, face detection, speech processing,
etc. Sometimes work in this area is also referred to as
``neuromorphic,'' since it can trace its roots to Biology as well. In
what follows, we do not examine these more recent ``deep neural
networks,'' since while the neurons have Biological connections, the
communication structure they use is engineered for functionality, without any attempt to mimic Biology.
% \begin{figure}
%     \centering
%     \includegraphics[scale=0.65]{figs/neural_computing.pdf}
%     \caption{Neural computing structures~(b) vs von Neumman-based computer architectures~(a). unlike conventional computer systems, in neural computing paradigms, memory is distributed among computing units while in conventional computers,   }
%     \label{fig:neural_computing}
% \end{figure}
Neuromorphic systems can be modeled entirely in software, and several
projects have developed tools to model Biological neurons, synapses,
axons, and dendrites at varying levels of detail~\cite{neuro_sw1,neuro_sw2,neuro_sw3}. However, such
systems are very slow. Even a simplified model of Biological
neural networks implemented on a massively parallel machine
ran 10$\times$ slower than real-time, and consumed 655\,kW of
power~\cite{bluebrain}. A Biological system of the same complexity takes five orders of magnitude less power while operating in real-time. 
\revised{connecting}{Hence, in order to explore the potential of Biologically-inspired computing systems for solving real-world problems, new types of energy-efficient devices and hardware architectures with high scalability factors are required. A domain where such systems could have a large impact is the Internet of Things (IoT), where small form-factor battery operated devices could be enhanced with capabilities akin to the sensory system (e.g. vision or speech recognition) of Biological systems.}

Hardware neuromorphic computing systems~\cite{Merolla_etal14a, Moradi_Indiveri11, Moradi_Indiveri14,Benjamin_etal14,Furber_etal14,loihi_micro,Park_etal16, caviar,jetc15} contain structures that
model neurons, synapses, and the connectivity between neurons and
synapses. \revised{re-organization}{Current large-scale systems have been implemented using conventional CMOS technology, and there has been a large body of work in recent years on novel memory devices for reducing the power consumption in neuromorphic systems. We begin with a brief overview of this body of work (Section~2), summarizing the focus of researchers developing novel devices for neuromorphic system. We then present the problem of implementing connectivity in neuromorphic systems and discusses some of the issues
involved in building large-scale neuromorphic hardware that can model millions of neurons (Section~3). We argue that the cost of communication in such systems is extremely high due to the complexity of supporting flexible high-fanout connectivity. This is consistent in what has been observed in existing large-scale neuromorphic systems (Section~4). 
We discuss what new materials/device technologies might be able to do to
alleviate this problem (Section 5), and provide possible directions
for future research in this area.}

\revised{thesis}{Our primary thesis is that almost all the recent work in memory technologies and materials/device design for neuromorphic systems cannot have a major impact on the {\it system-level\/} power consumption of large-scale neuromorphic systems without new techniques being developed to address the communication problem. When examining the complete system, it is quite clear that the focus of device researchers has been on only a portion of the total power requirement; a non-trivial fraction of the total power budget remains unaddressed,  limiting the benefits of the proposed device technologies in isolation.}

\section{Resistive memory technologies for neuromorphic systems}
\begin{figure}
\centering
\includegraphics[scale=0.45]{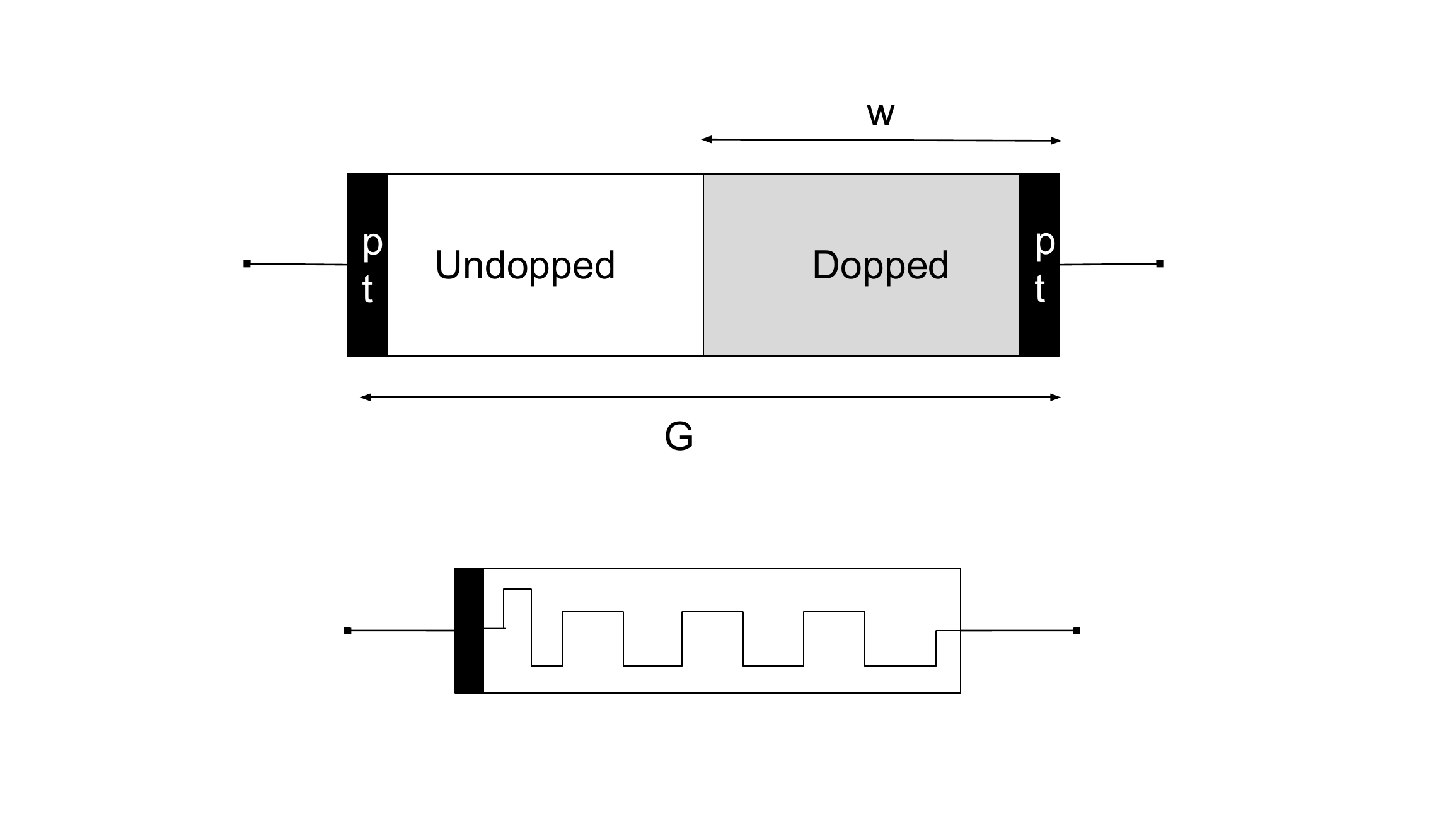}
\caption{Generic Memristor structure and its symbol~\cite{springerBook2013}.}
\label{fig:hpmem}
\end{figure}

Most prominent large-scale neuromorphic projects have employed commercial CMOS technology for implementing neural principles. In recent years, there has been growing research efforts to utilize emergent memory and device technologies for scaling up neuromorphic hardware systems. In neuromorphic engineering, most such technologies are primary considered for implementing synaptic and learning models. However, there is a potential to utilize nanoscale emergent devices to overcome the interconnects limitations~\cite{dally_interconnection} in 2D\,/3D VLSI integrations. In recent neuromorphic designs~\cite{dynaps,loihi_micro,Merolla_etal14a}, the communication and standby power has become a significant portion of total power consumption. In this section we review most relevant research articles that have investigated the potential of two-terminal emergent devices in improving the performance of CMOS neuromorphic computing systems.  

%\subsection{Two-terminal nanoscale non-volatile memory devices }
Multiple emergent memory devices have been recently proposed to overcome the limitations of the conventional CMOS memories~\cite{Indiveri_memory_limitation} in embedded systems. Memory resistive~(memristor) elements, oxide-based resistive random access memory~(oxRAM), conductive-bridging random access memory~(CBRAM), phase-change memory~(PCRAM) are examples of the emergent memory devices that are explored for integrating with CMOS-based neuromorphic circuits. The characterizations of these memory devices varies depending on the switching materials. In order to incorporate the emergent devices with CMOS-based neuromorphic systems, certain design constraints need to be satisfied~\cite{emergent_neuro_consideration}. The main idea has been to use such memory elements for implementing synaptic computation in a more compact and energy-efficient way in compare to their CMOS counterparts. Given the large number of synapses in neuromorphic networks, more optimal synapse circuits leads to significant improvements in overall performance. 
%Non-volatile memristors define the relation between electric charge and magnetic flux.
While the term memristor was introduced in 1971~\cite{chua}, the first experimental prototypes were only recently demonstrated by HP Labs~\cite{HP}.
\begin{figure}
    \centering
    \includegraphics[scale=0.35]{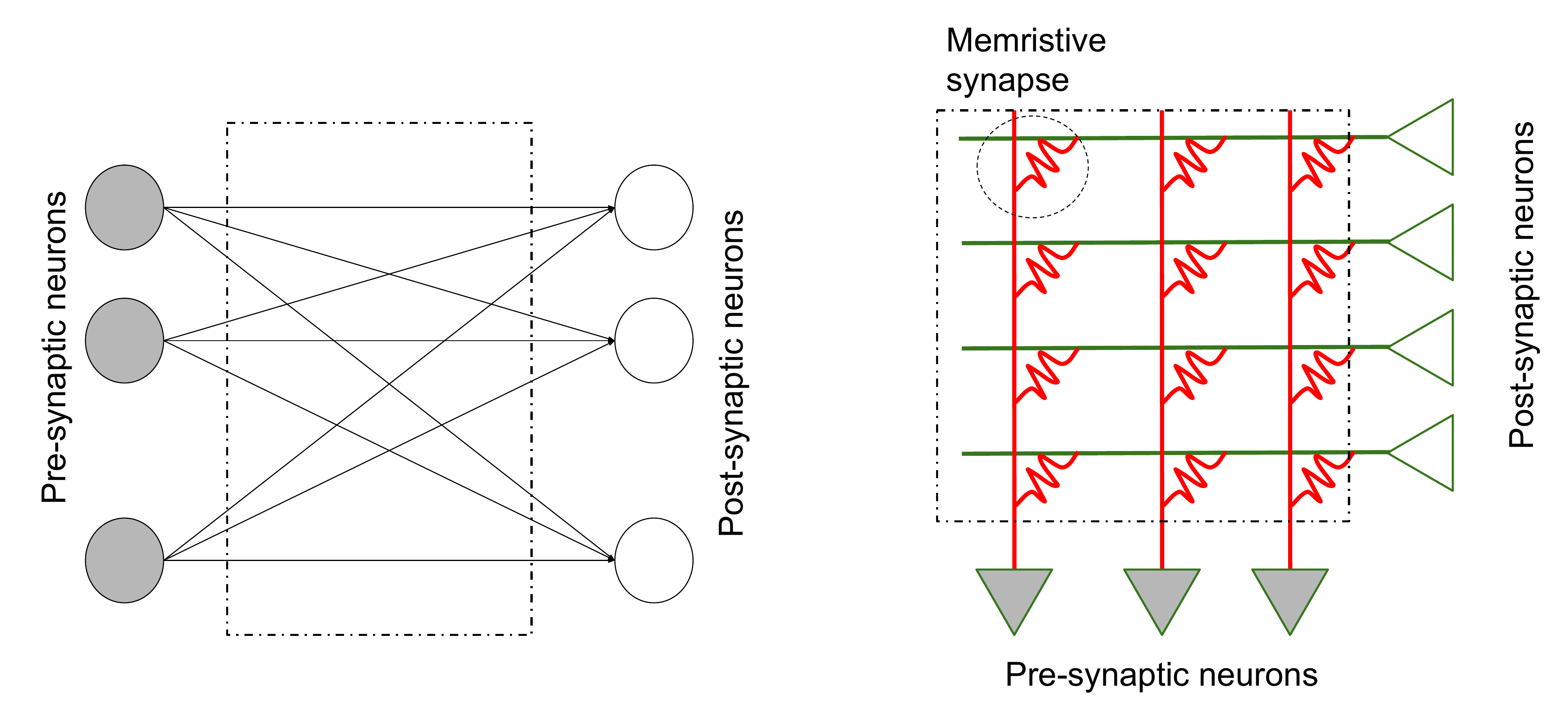}
    \caption{Left: 2-layer feedforward neural network. Right: memristor-based neuromorphic cross-bar module. In such configuration, a memristor device is used to implement synapse units, axons~(in red) and dendrites~(in green). }
    \label{fig:memrsitor}
\end{figure}

\revised{memristor}{Fig.\ref{fig:hpmem} illustrate the generic structure of memristive devices. The memristor can be considered as two regions in series, doped and undoped. Ideally, the resistance difference between doped and undoped region is very high and the overall device resistance is a function of doping ratio(\textit{w}/G), where \textit{w} the length of the doped region, and $G$ is the thickness of the memristor.  The size of these regions can be changed depending on amount and direction of the charge that flow through the device. For instance in a voltage-controlled, by applying a positive voltage increases the thickness of doped region increases until $\textit{w}=1$, and the resistance will be very low or device is in ON state.}
% By tuning oxidation~\cite{customized_memristor}, the memristor devices can be configured to operate in a bi-stable memory state---\emph{i.e.} a low resistance state~(LRS) or a high resistance state~(HRS). 
Ideally, memristors can be programmed to represent analog memory or various intermediate resistance states between HRS and LRS~\cite{memristor_1t1r}. Because of their programmable resistive states and small footprint~(in order of a few\,nm), memristors are promising candidates for efficiently implementing the synaptic connection between pre-synaptic and post-synaptic neurons. There has been significant activity dedicated to developing memristive-based neuromorphic systems~\cite{afifi_09, hu2014memristor, stdp_memrisotr, Indiveri_memory_limitation}. 
A variety of oxide-based materials~(e.g. $Pr_{1-x}Ca_{x}MnO_{3}$,$WO_{x}$, $AlO_{x}$, $HfO_{x}$, $TaO_{x}$, $TiO_{x}$) have been investigated to implement the synaptic array in neuromorphic systems~\cite{ReRAM_analog,ReRAM_lv,oxReRAM_survey}, for applications such as pattern recognition~\cite{ReRAM_pattern}, face detection~\cite{ReRAM_face}, learning algorithms~\cite{ReRAM_learning}. Further details on memristive synapse implementations is beyond the scope of this paper. However readers can refer to Hong et al.~\cite{oxReRAM_survey} for a comprehensive survey of oxide-based memristor devices and their benchmarking metrics for neuromorphic computation.

\revised{memristor-arch}{Crossbar-based structures are typically used to implement synaptic arrays. In crossbar schemes, the memristors are placed on the cross-point between vertical and horizontal wires. However by increasing the size of crossbar array the power consumption dramatically increases. This happens due to sneak current path of memristive-based memory elements as well as other alternate current path. It is therefore memristors are typically placed on the grid point in series with an access transistor to limit the sneak current. Also referred to as 1T1R~(one-transistor-one-resistor), such scheme leads to larger cell size~e.g.~$20F^{2}$, and lowers the memory density. 1D1R~(one-diode-one-resistor) and 0T1R~\cite{isqed2017} are among other solutions that have been investigated for limiting the sneak current issues~\cite{1d1r}.}

Hybrid CMOS-memristor based architectures have been proposed based on the cross-net structure~\cite{crossnet}, where the idea has been to use a grid of synaptic memristive devices on the top of CMOS implemented neuronal arrays~(see fig.~\ref{fig:memrsitor}). Successful implementations of such a scheme can help to significantly scale the number neurons and synapses on a single die compared to standard CMOS technology. The memristor devices have also been utilized to implement training algorithms, for instance back-propagation~\cite{hassan} and biologically-inspired learning rules e.g. Spiking-timing Dependent Plasticity~(STDP). Furthermore, the potential of all-memristive neuromorphic computing system has also been examined~\cite{all_memristive}. CBRAM~\cite{cbram_binary1, cbram_binary2} and oxRAM~\cite{oxram_cbram} are examples of other nanoscale devices that have been studied for implementing synapse circuits. Their structure is based on Metal-Insulator-Metal~(MIM) configuration, and are typically used as binary memory elements. While it is possible to program these devices with multi-bit resolutions, it is often achieved at the cost of a higher variability and larger control circuits~\cite{oxram_cost}.  These devices are typically fabricated using different materials and techniques~\cite{zheng_15}, and therefore, depending on configuration, represent different electrical characteristics.

Although the efficiency of memristor devices has been successfully demonstrated for single device or a small array of synapses~\cite{hu2014memristor}, the potential of large-scale CMOS-memristor based design are still being explored in academia. The device reliability, manufacturability, and yield issues are major obstacles in implementing a large synaptic matrix based on the emergent nanoscale memories. Combinations of different switching materials are being investigated in order to overcome these limitations in large-scale integrations of neuromorphic CMOS and memristors. If successful, it would further reduce the power requirements for performing computation. In the next section, we discuss the communication requirements of next-generation scalable low-power neuromorphic processors, and investigate the impact of communication and computation in overall power consumption.  

\section{Communication requirements}
\label{sec:communication}
\begin{figure}
\begin{center}
\includegraphics[scale=0.4]{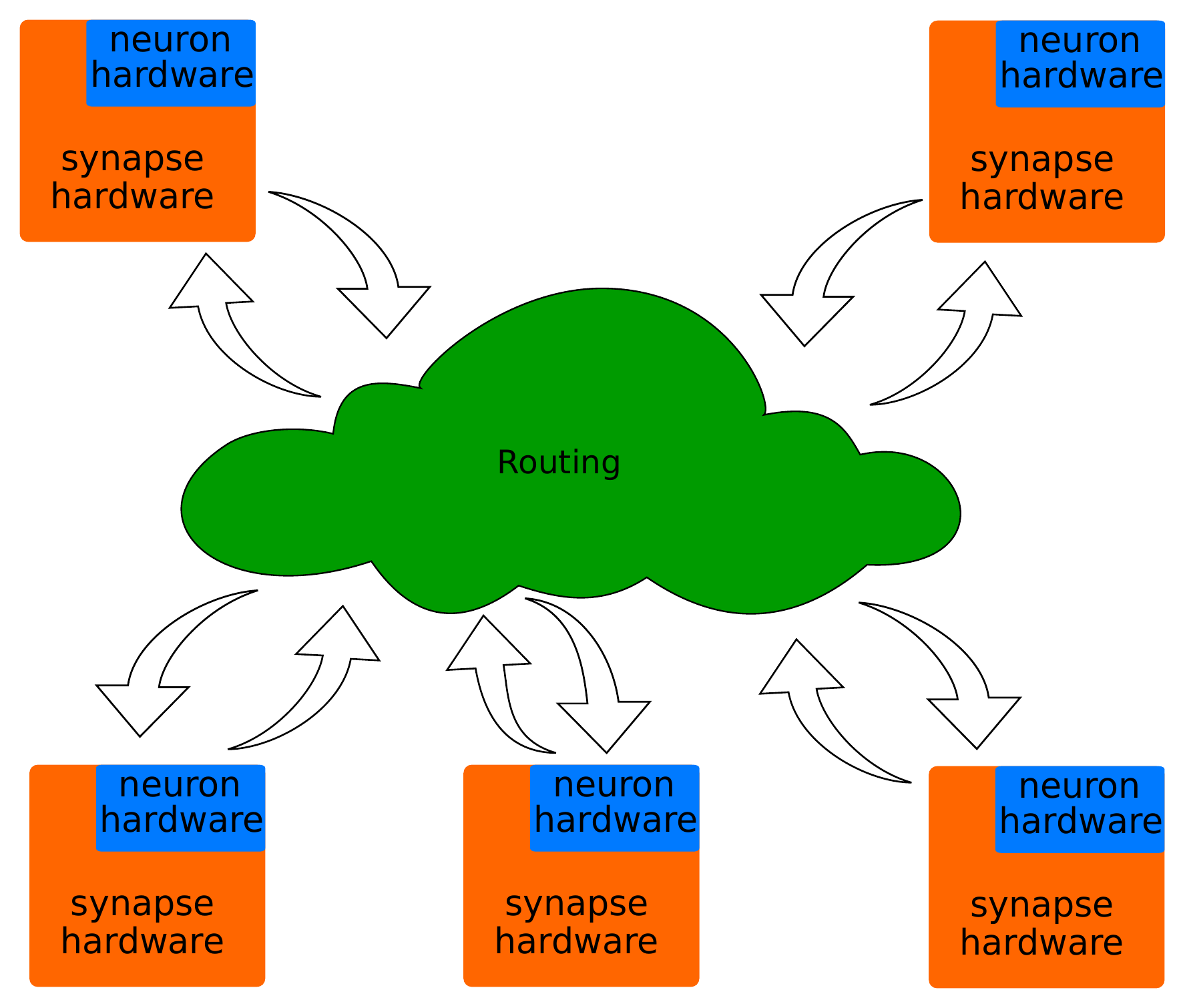}
\caption{Generic parallel neuromorphic hardware architecture.}
\label{fig:sketch}
\end{center}
\end{figure}
% \begin{figure}
% \centering
% \begin{subfigure}{.5\textwidth}
%   \centering
% \includegraphics[scale=0.42]{figs/generic_arch.pdf}
%   \caption{}
%   \label{fig:sketch}
% \end{subfigure}%
% \begin{subfigure}{.5\textwidth}
%   \centering
% \includegraphics[scale=0.7]{figs/BW_review.pdf}
%   \caption{}
%   \label{fig:BW_review}
% \end{subfigure}
% \caption{(a) Generic parallel neuromorphic hardware architecture. (b) Bandwidth and latency requirements in neuromorphic hardware. Assuming that the bisection bandwidth is $B$ and the spike latency of $l$, the network can transfer $B$ spikes in one time unit. If the delay per spike route is $o$, then $(B)\times~o = 1$ or $o = \frac{1}{B}$.}
% \label{fig:test}
% \end{figure}
A diverse variety of spike-based neuromorphic computing systems have been developed and implemented in silicon over the past three decades. There are many differences in the implementation details of each system, based on the goals of the implementation and the state of silicon technology at the time the system was created. For instance early implementations were dominated by analog circuit design, but completely or mostly digital implementations can be seen in recent years due to the scaling of CMOS device technology from a feature size of 3$\mu$m in the mid 1980s to 14nm and below today. Some systems aimed to faithfully mimic Biology---which included operating at Biological time constants. Others were aimed at accelerated modeling, with time constants many orders of magnitude faster than Biology. These choices have cascading implications that change the overall architecture of the neuromorphic system. In particular, communication requirements can change dramatically depending on the implementation goals.

Despite their many differences, neuromorphic systems all attempt to emulate Biology. This means that no matter how various aspects of Biology are modeled, the hardware representing neurons and synapses must logically operate in a massively parallel fashion. Hence, it is possible to view {\em every\/} large-scale neuromorphic hardware platform as a parallel collection of neurons and synapses with some communication infrastructure. Furthermore, it is widely accepted that neurons and synapses have dense local connectivity and relatively sparse global connectivity. Hence, every platform consists of a collection of {\em clusters\/} of neurons and synapses that support efficient local communication within the cluster combined with a global inter-cluster communication network (Fig.~\ref{fig:sketch}).

The flexibility of a routing architecture determines the connectivity between neurons and synapses that is natively supported by the hardware. It should come as no surprise that there is a high cost to be paid for supporting highly flexible connectivity, whereas systems that are more restrictive can be made more efficient. The cost of flexibility is higher design complexity and more hardware resources including on-chip memory. In this section, we examine the communication requirements for a generic spiking neural network and possible trade-offs between the system flexibility and its complexity. 

Consider a system that models $N$ neurons and $S$ synapses per neuron. ($N\approx 10^{14}$ for the mammalian brain, and $S\approx 10^4$.)  Since the average number of incoming connections to a neuron from other neurons is $S$, it follows that the average outgoing connections from a neuron is also $S$ (with some exceptions for external inputs and outputs). A neuromorphic communication network has to support an average fan-out of roughly $10^4$ in silicon. Static usage of wiring resources for this purpose is infeasible, since VLSI systems have always been wiring limited even when the average signal fanout is in the range of three to four destinations. Therefore, packet switching is the architecture of choice for global communication in neuromorphic systems. The speed of silicon device switching is used to time-multiplex wires to provide support for global communication. The {\em address-event representation\/}~(AER) is most commonly used communication protocol in neuromorphic systems. In the AER protocol, each neuron is given a unique address and upon generating a spike, the addresses packet is routed to the destinations of the source neuron. Hence, the global inter-cluster communication network is a packet-switched communication network, and its single-chip component is a network-on-a-chip (NoC).

\subsection{Routing memory}
Assume a generic spiking neural network with $N$ neurons and $S$ synapses per neuron. This results in total $N\times S$ destinations. In conventional routing schemes used by parallel machines, each destination would be assigned a unique address encoded with $\lg(NS)$ bits. In order to support arbitrary connectivity, $NS\lg(NS)$ bits would have to be stored---which corresponds to $\mathord{\approx}38.7$GB of storage for a million neuron system ($40.5$KB per neuron), which would dominate the silicon area needed for the system. Many techniques have been introduced to reduce the storage required. One approach groups the synapses for a neuron by their type e.g. inhibitory vs excitatory, fast vs slow, etc. Modeling the computation performed by the synapse as a linear filter permits the per-synapse state to be grouped by synapse {\em type\/} using the principle of superposition~\cite{syn_superposable}. If there are $k$ types of synapses, the storage requirements would be $NS\lg(kN)$. However, even for small $k$ (e.g. $k=4$), this only reduces the storage by a factor of $1.5$. For this reason, state of the art neuromorphic systems typically limit the flexibility of supported connectivity in order to limit the memory needed to store connectivity information.

\subsection{Latency constraints}
In neuromorphic hardware systems, spikes are represented by binary addresses sent in the form of packets across the network. The packets carry the spikes' timing~(implicitly) and contain the routing information to create the virtual connection(s) between a source node and its destination(s).  The packets may include other information such as axonal delays or synaptic weights. In terms of the size, the spike packets in neuromorphic hardware are typically smaller of those of conventional communication networks, consisting of little more than the packet header that stores routing information. Since neuromorphic systems represent information in the {\em timing\/} of spikes, it is important that the communication network preserve the delivery time of a spike relative to the time it was generated. In other words, the communication network must have predictable latency relative to the time scale of the computation. In what follows, we assume a packet switched network consisting of a collection of routers interconnected in some pre-defined topology~\cite{dally_interconnection}.
\begin{figure}
\begin{center}
\includegraphics[scale=0.7]{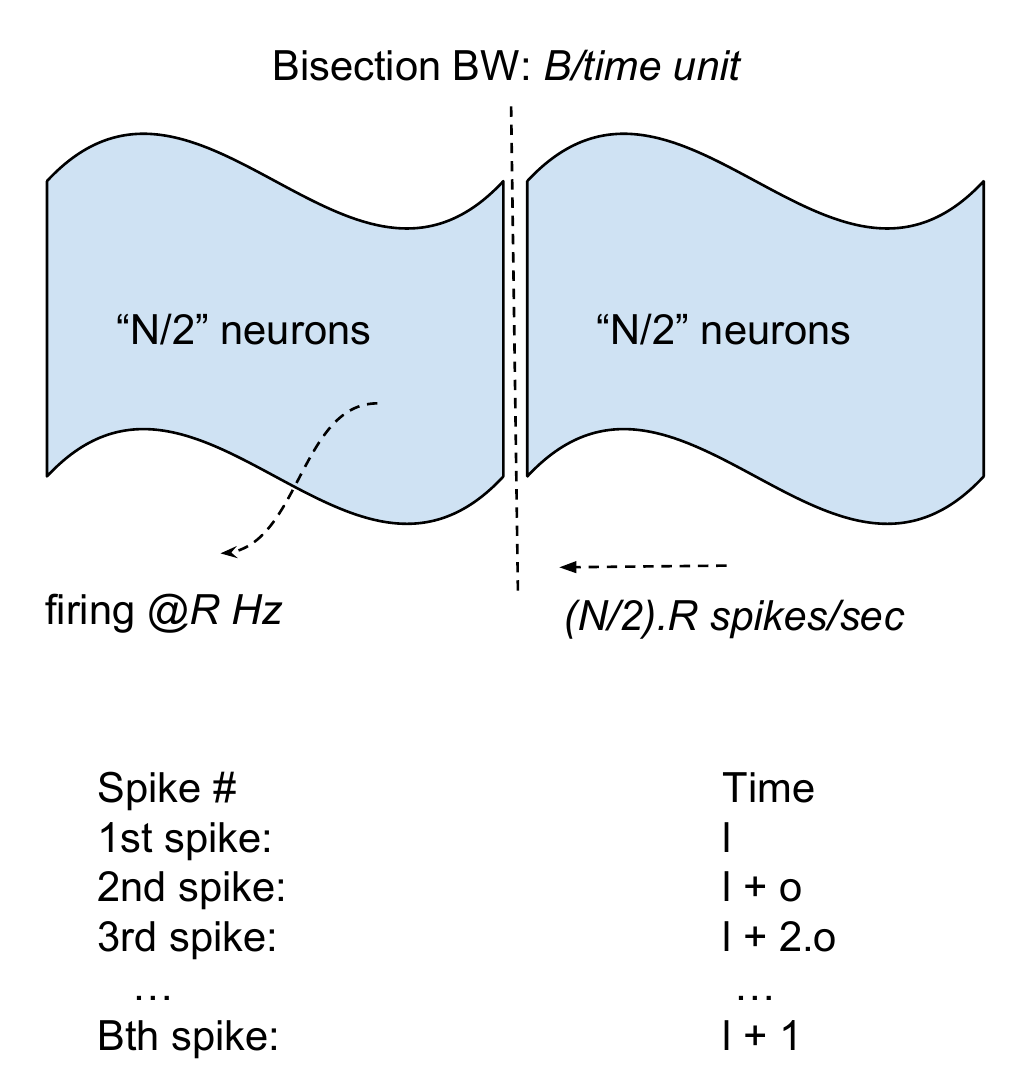}
\caption{Bandwidth and latency requirements in neuromorphic hardware. Assuming bisection bandwidth $B$ and the spike latency of $l$, the network can transfer $B$ spikes in one time unit. If the delay per spike route is $o$, then $(B)\times~o = 1$ or $o = \frac{1}{B}$.}
\label{fig:BW_review}
\end{center}
\end{figure}
The bandwidth and timing requirements are the important factors in the design of large-scale neuromorphic hardware systems. Fig.~\ref{fig:BW_review} illustrates a sketch of network with $N$ neurons with firing rate of $R$ ($R$ is typically $10$Hz). Consider a very simple network model, where the network has a bisection bandwidth $B$ spikes/s, spike latency of $l$, and mean router link occupancy per spike of $o$. Since the bandwidth of a link is determined by the rate at which it can process spikes, the link bandwidth is $1/o$ spikes/s. Such a network requires $l$ fraction of time unit to route the first spike, $l+o$ for the second spike and so on for spikes sharing a link. The number of links between the two halves of the network in Fig.~\ref{fig:BW_review} will be $C=B\times o$. If all $\frac{N}{2}$ neurons in one half of the network generate spikes in a synchronized fashion, and the spikes have to be delivered to the other half, the network can support the traffic if $B \ge \frac{NR}{2}$. However, the number of links and not just the bandwidth matters since that impacts the packet latency. Since there are $C$ links, even under uniform distribution of traffic, the last group of packets will have a latency of $l+(\frac{NR}{2C}-1)o = l + (\frac{NR}{2B} - \frac{C}{B})$ while the first set of packets will have a latency of $l$. Hence, the uncertainty in spike arrival will be in the range of $[0,\frac{NR}{2B}-\frac{C}{B}]$ relative to when the spikes were generated. Note that for a fixed bisection bandwidth, it is better to have more parallel links (higher $C$) with lower bandwidth per link if the traffic can be distributed across them. (Note that the Biological extreme of dedicated wiring minimizes this quantity, but is infeasible in current silicon technology.)

For a firing rate of $R=10$Hz, Biological models typically care about temporal precision in the $0.1$ms range~\cite{temporal_precision} (i.e. $1/({10^3\times R})$). Using this rule of thumb to bound the timing uncertainty, we have that 
$\frac{NR}{2B}-\frac{C}{B} \le \frac{1}{10^3\times R}$, i.e. $B \ge 10^3(NR^2/2-C)$---a significantly stricter requirement than $B \ge NR/2$ for conventional networks where the packet timing does not carry information. For accelerated network modeling (i.e. values of $R$ significantly faster than real-time), there is ``double penalty'' for the speedup: one from simply a larger volume of  traffic, and a second from a much stricter latency delivery constraint. For $N=10^6$ and $R=10$Hz, even with $\approx 1000$ links between two halves of the network, this comes to $B \approx 50$Gspikes/s rather than $5$Mspikes/s for traditional networks. Note that both numbers scale down once we assume significant communication {\em locality\/}, i.e. if we assume that $\alpha N/2$ rather than $N/2$ neurons generate spikes that have to cross over to the other half of the network. Also, the bandwidth requirement can be reduced by relaxing the temporal precision used by the system---going from a $0.1$ms tolerance to a $1$ms tolerance would cut the bandwidth requirement by a factor of ten.
This is why most state-of-the-art neuromorphic systems have networks that at first glance look highly over-designed and underutilized---the latency constraint compounds the network bandwidth requirement.

\subsection{Topology and router design}
The NoC architecture is specified by its topology and routing strategy. Different topologies have been adopted to support higher throughput demands over time. One-dimensional, two-dimensional, and hierarchical networks are examples of NoC topologies that are utilized by the neuromorphic community. The link bandwidth between routers in a NoC can also be estimated using an analysis similar to the one above. Assuming that a cluster of neurons has on average $N_c$ neurons that have to communicate with neurons in a different cluster, if $r$ is the average degree of a router and the average number of links traversed by a spike is $d$, then each link will have to support traffic that is approximately $N_cRd/r$. Using an argument similar to the one above, the latency uncertainty of a link is $[0,(\frac{N_cRd}{r}-1)o]$, where $1/o$ is the link bandwidth. Hence, the link bandwidth requirement is $1/o\ge 10^3d(N_c R^2 d/r-1)$. An analysis without latency constraints would result in $1/o\ge (N_c Rd/r-1)$. Hence, the network has to be designed to  support much higher per link bandwidth to preserve spike timing, even though only a small fraction of the bandwidth is utilized in practice.

\begin{figure}
%\centering
\begin{subfigure}{.5\textwidth}
% \centering
\includegraphics[scale=0.41]{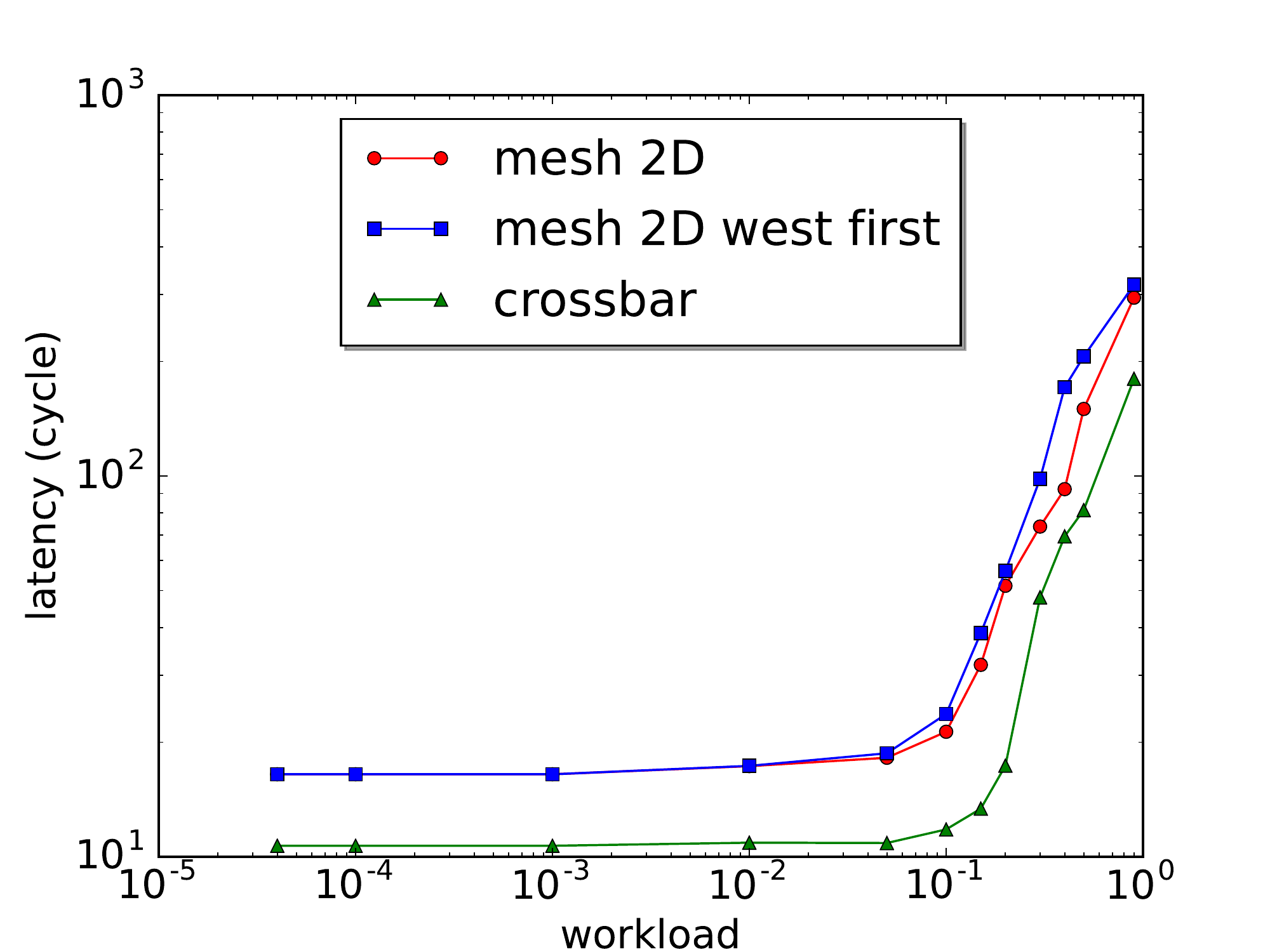}
  \caption{Network Latency.}
  \label{fig:sub11}
\end{subfigure}%
\begin{subfigure}{.5\textwidth}
  \centering
\includegraphics[scale=0.41]{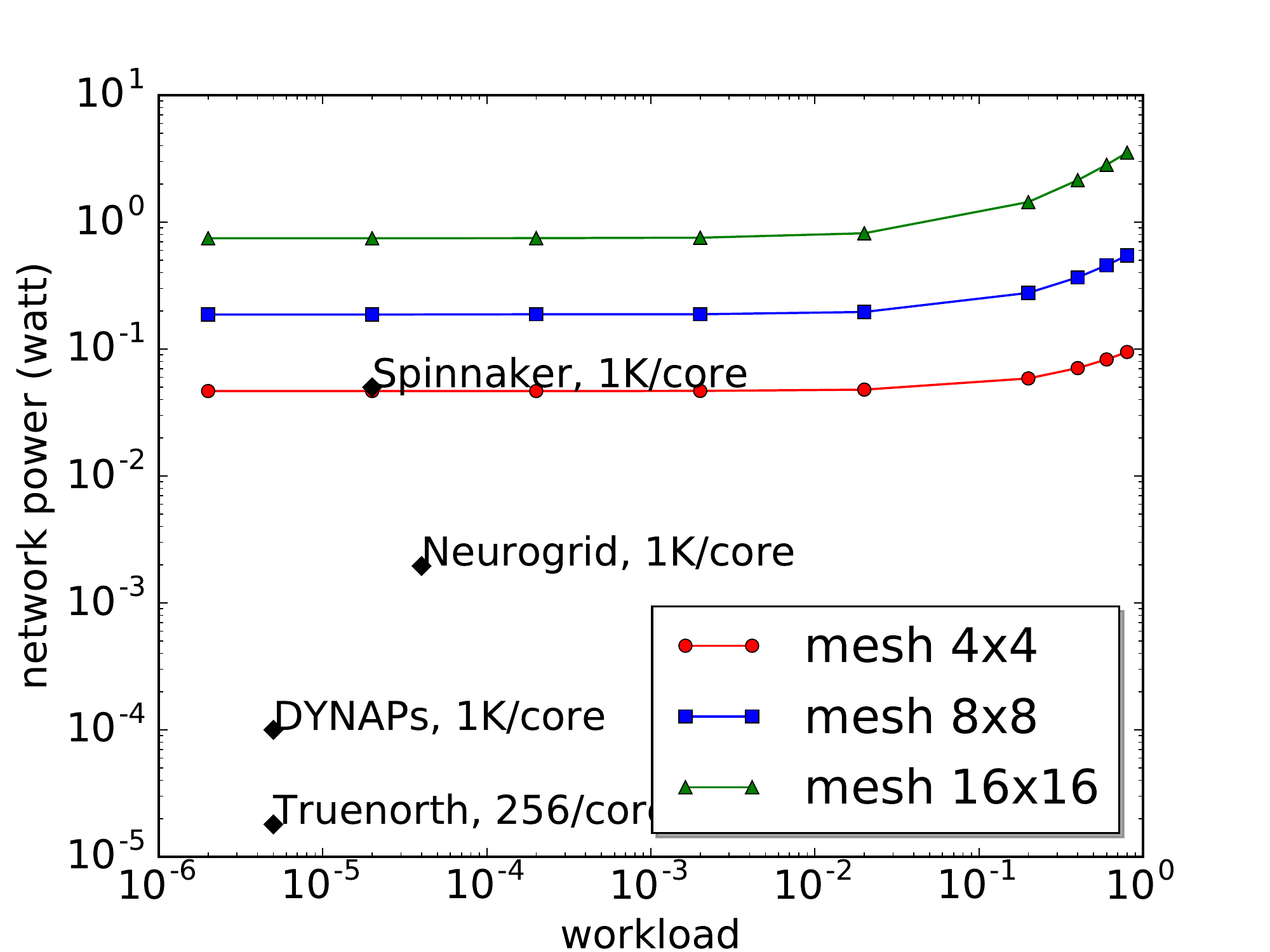}
  \caption{Network power estimates.}
  \label{fig:sub22}
\end{subfigure}
\caption{ (a): The network latency versus network workload for different network \& routing topologies simulated using~\cite{garnet}, frequency 1\,GHz and synthetic traffic patterns. (b): Power consumption for different 2D-mesh network sizes, estimated using the DSENT tool~\cite{dsent}, and the existing neural cores: Spinnkaer~\cite{Furber_etal14}, Neurogrid~\cite{Benjamin_etal14}, DYNAPs~\cite{dynaps} and Truenorth~\cite{Merolla_etal14a}.} 
\label{fig:routing}
\end{figure}
A major difference between the neuromorphic and conventional NoCs is the amount of network traffic generated by a node in the network. In contrast to conventional applications of NoCs, the computing nodes~(neurons) in neuromorphic systems run at very low frequencies~(in order of Hz), and so drastically lower traffic is injected into the routing fabric. 
Fig.~\ref{fig:sub11} shows the average latency for different networks with high workloads as simulated by a NoC simulator used for computer architecture studies~\cite{garnet}. As the figure suggests, the latency in the networks with high workload varies significantly by network topology and routing algorithm in addition to the network load. However, as we have argued above, the link bandwidth has to be over-provisioned relative to conventional networks in order to keep the spike delivery uncertainty small. Hence, as far as the network is concerned, we are operating in a regime with very low workload compared to the peak feasible bandwidth. As can be seen (Fig.~\ref{fig:sub11}), in networks with very low workload~(below 0.01 spike/neuron/cycle), the routing methodology and the network load does not significantly impact the network latency. Given the cycle time of NoC digital circuits~(at most a few nanoseconds) and the low neuronal firing rates~(in order of a few\,Hz), most neuromorphic NoC workloads fall well below 0.001 spikes/neuron/cycle. This characteristic means that the network topology and routing algorithm is not as important in systems operating at Biological time scales compared to conventional NoCs.

Another differentiating feature is that the power consumption of a neuromorphic core is usually substantially lower than that of conventional compute cores like microprocessors that are traditionally used with NoC fabrics. For example, in both the TrueNorth~\cite{Merolla_etal14a} and DYNAPs~\cite{dynaps} projects, the neural cores are custom designed and require a fraction of the power consumption of a standard low power microprocessor. So, for neuromorphic systems, the power consumption of the routing fabric can be a significant portion of the total system power. Fig.~\ref{fig:sub22} shows the estimated network power used by an architectural NoC simulator~\cite{dsent} that is used to study on-chip networks. Apart from the SpiNNaker project (which consists of a large number of ARM cores as compute nodes), other neuromorphic systems have networks that consume orders of magnitude less power than those modeled by the NoC simulator because they are much more parsimonious in their hardware usage.
We note that the estimated power of different neuromorphic networks in Fig.~\ref{fig:sub22} are not intended for direct comparison with each other, as they represent different choices in terms of network flexibility; we simply note that reducing the network power consumption is an important aspect of neuromorphic hardware design. In the light of new memory technologies, the power gap between the computation and communication in custom-designed neuromorphic systems is likely to increase, making the network power consumption even more important in future neuromorphic systems.

\subsection{Existing Large-scale Neuromorphic CMOS Architectures}
\label{sec:cmos}
We briefly review major existing neuromorphic hardware systems and their design choices in addressing the design scalability challenges. As will be evident, each system is organized as a cluster of neurons and synapses with a global inter-cluster communication network. Table~\ref{table:np_comparisons} shows examples of representative neuromorphic hardware systems that have been developed by academic and industrial research teams.

The system architecture developed during the SyNAPSE project, called
TrueNorth~\cite{Merolla_etal14a}, is the largest single chip built
within the neuromorphic community with 1\,million neurons,
256\,million synapses, and a network of 4096 neurosynaptic cores. Each
core contains 256 neurons and 256$\times$256 synapses using SRAM-based
cross-bar arrays. The neurons and synapses implemented in the
``neurosynaptic core'' are designed to have behavior that matches a
deterministic software model for the system. The global routing topology is a two-dimensional mesh architecture. Thanks to the power
efficiency of asynchronous circuits and the use of low-leakage CMOS
process, the power figure reported for TrueNorth is its most
distinguishable feature compared to other neuromorphic hardware
efforts. The chip consumes only 72\,mW---significantly lower than
other neuromorphic substrates~\cite{Merolla_etal14a}.

The Loihi chip~\cite{loihi_micro}, designed by Intel, is a many core neuromorphic architecture that comprises 128 neural cores, three embedded x86 processors and asynchronous 2D mesh network-on-chip routing fabric for connecting the neural cores. Each neuromorphic core is equipped with a programmable learning engine to implement different training algorithms. The learning capability and more flexible routing network distinguish Loihi from prior large-scale custom-designed neuromorphic systems. The communication architecture is also mesh based. Loihi is a fully digital architecture implemented in 14\,nm CMOS process, and its  routing network is designed using asynchronous circuits.    

The Neurogrid project consists of a system that comprises almost entirely of custom mixed-signal hardware for modeling biological neurons and synapses. The core hardware element in Neurogrid is the Neurocore chip, which is a custom ASIC that uses analog VLSI to implement neurons and synapses, and digital asynchronous VLSI to implement spike-based communication. The chip was fabricated in a 180nm process technology, and contains a 256$\times$256 array of neurons whose core functionality is implemented with analog circuits. Each neuron is
implemented using custom analog circuitry that directly implements a a quadratic integrate-and-fire (QIF) model, and is combined with four types of synapse circuits~\cite{Benjamin_etal14}. The routing architecture uses a tree topology, with hardware support for multicasting.
\begin{table}[btp]
%\footnotesize
%  \vspace{10pt}
%  \centering
\resizebox{0.97\columnwidth}{!}{%
  \begin{tabularx}{1.35\textwidth}{ l c c c c  }  
    \toprule
    \textbf{} & 
    \textbf{Truenorth~\cite{Merolla_etal14a}} & 
    \textbf{Spinnaker~\cite{Furber_etal14}} & 
%    \textbf{HiAER~\cite{Park_etal16}} & 
    \textbf{Neurogrid~\cite{Benjamin_etal14}} & 
    \textbf{Loihi~\cite{loihi_micro}}\\ %[0.5ex] 
    \midrule
    Solution  & Chip  & Board &  Board & Chip\\ 
    Technology & 28\,nm  & 130\,nm &  180\,nm & 14\,nm  \\ 
    Comput. core & custom-digital & Commercial-digital & Custom-Mixed & Custom-digital \\
    Processing type & Time multiplexing & Time multiplexing & Parallel & Parallel \\
    No. of neurons & 1\,M & 846\,K & 1\,M & 131\,K\\
    On-chip Learning & No & Yes &  No & Yes\\
    Power consumption  & 72\,mW  & $\sim$75\,W & 3\,W & -- \\ 
    Computation power$\sim$  & 30\,$\%$  & 52\,$\%$ & -- & --\\ 
    Communication Power$\sim$  & 10\,$\%$  & 12\,$\%$ & --  & --\\ 
    Static Power$\sim$  & 60\,$\%$  & 36\,$\%$ & -- &-- \\ 
    \bottomrule
  \end{tabularx}%
  }
  \caption{The specifications of the major neuromorphic hardware projects. The power breakdown numbers are provided in the reference articles, except for Neurogrid and Loihi chips.}  
  \label{table:np_comparisons} 
\end{table}

The SpiNNaker (Spiking Neural Network Architecture) project~\cite{Furber_etal14} at the
University of Manchester has taken the most ``general-purpose'' approach to the design of large-scale neuromorphic systems. The core hardware element for the neuromorphic system is a custom-designed ASIC called the SpiNNaker chip. This chip includes eighteen ARM processor nodes~(the ARM968 core available from ARM Ltd, one of the project’s
industrial partners), and a specially designed router for communication between SpiNNaker chips~\cite{painkras2013spinnaker}. The routing topology is a two-dimensional torus, with additional diagonal connections. A complete SpiNNaker board contains 47 of these chips, and the goal is to assemble 1200 of these boards. A full SpiNNaker system of this size would consume about 90\,KW.
%The HiAER (for hierarchical address-event representation) design~\cite{Park_etal16} combines custom neurons and synapse modeling hardware with
%programmable routing. The system consists of two types of components: the IFAT (integrate-and-fire array tranceiver) board, and the routing chip. Each IFAT system implements an array of neurons and their associated synapses, and has routing resources for spike communication. Sets of IFAT/router nodes are connected in a linear array, and the edge of each array is responsible for communication to other arrays. These edge nodes are also organized in linear arrays, and so the entire system can be viewed as a hierarchy of arrays.
%\subsection{Impact of Network topology and routing methods in large-scale neuromorphic electronic systems}

\subsection{Summary}

\revised{summary-cmos}{When examining a complete neuromorphic system, a few items become quite clear. First, the static power consumption is a large fraction of the total power budget. This is because neuromorphic systems operate at very slow timescales compared to the capabilities of modern CMOS devices. The traditional approach taken by conventional digital systems---namely running computation at a very high throughput, which amortizes the static power consumption---is not as easy to accomplish in the neuromorphic domain due to the ``double penalty'' imposed by strict latency constraints. Note that while our analytical modeling is quite simple, its simplicity is what makes it applicable to a large class of neuromormophic systems.
Second, traditional on-chip network architectures consume significantly more power than the custom-designed networks in neuromorphic chips. In spite of this, the network power is a double-digit percentage of the total system power.  This means that even if the rest of the system took zero power, the maximum power benefit would be limited by a factor of ten. Finally, the approach of using a large number of small clusters to achieve power efficiency means that any memory requirements for routing have to be distributed into a large number of small memories rather than a single monolithic memory. Hence, what becomes important is not how large a single memory bit is, but the {\it effective\/} density of a small memory array. The array efficiency of a memory array is the ratio of the area occupied by the storage elements (the bits) to the total area of the memory---which includes addressing as well as read/write circuitry. Even SRAMs (which arguably have the simplest external circuitry while having the largest storage element area) have array efficiencies as low as 50\%-80\% for sizes below 1Mb. A smaller storage element proposed by numerous researchers makes the array efficiency even worse when the memory size is fixed. This observation was validated in a design that used phase-change memory technology as the synaptic element~\cite{IBMPCRAMpaper}.}

\section{Less explored device technologies for neuromorphic hardware}

In this section, we present some less-explored device technologies that have been proposed to improve the performance of memory systems, that also have applicability in improving the scalability of neuromorphic systems.
% \begin{figure}
%     \centering
%     \includegraphics[scale=0.3]{figs/TSV.png}
%     \caption{Cross-section view of 3D integration of silicon dies with TSV technology. A TSV channels in 3D IC includes through silicon via, bumps, and the redistribution layers~(RDL). (figure from~\cite{tsv_freq} \copyright IEEE)}
%     \label{fig:tsv}
% \end{figure}
\subsection{Three-dimensional integrated circuits}

Three-dimensional (3D) integration has been explored as a way to improve memory system performance by vertical stacking of multiple memory device layers. This technology is mature enough so that it is commercially available, as it has demonstrable advantages in improving integration density of memory with logic. Also, since neuromorphic circuits operate with extremely low power budgets, one of the major concerns in 3D integration---the problem of heat dissipation---is alleviated.

Neuromorphic networks are typically composed of a massively parallel collection of neurons and synapses with large neuronal fan-outs. A vast majority of hardware resources in neuromorphic systems are therefore dedicated to the interconnections between the computing modules. Unlike biology, neuromorphic VLSI systems are implemented in 2D substrates. The large number of connections imposes serious design challenges for scaling up the size of neuromorphic networks in VLSI. 
Recent advances in semiconductor industry~\textit{e.g. flip-chip and Through-silicon-vias~(TSV)} offer vertical integrations of integrated circuits. Such integrations potentially would lead to significant performance improvements in implementing highly interconnected networks, in comparison to their planar counterparts. TSV technology~\cite{tsv_design,tsv_design1} is one of the most promising solutions for stacking of multiple dies in a small package and for memory integration with conventional CMOS. %Fig.~\ref{fig:tsv} illustrates a cross-section view of TSV channels in vertical integration of silicons dies. In addition to through silicon via, a channel includes bumps and redistribution layers~(RDL)~\cite{tsv_freq}. 
3D TSV-based ICs are used in stacked memory and processing systems~\cite{tsv_mem_proc,tsv_mem_proc1}, CMOS\&MEMS image sensors~\cite{tsv_sensor1}, IoT Socs~\cite{tsv_soc}, and 3D FPGAs~\cite{tsv_fpga}. Utilizing TSV in vertical integrations reduces wire complexity, the 2D die size, and the interconnection delays in high density interconnect networks. 
However, such integrations face serious technological challenges from design and physical stand point. A major drawback in design of 3D TSV ICs is the lack of commercial EDA tool support, and accurate electrical models that could capture the physical properties of the TSV connections integrated with CMOS circuits~\cite{tsv_models}. IC design tools equipped with detailed TSV characterization models can greatly help in investigating the full potential of this technology across different application domains. Depending on operating frequency, TSV links can have resistive or capacitive behavior. A detailed model and frequency response of TSV channels are presented in~\cite{tsv_freq}. Thermal power and cross-talk issues are other important factors to be considered in utilizing TSV connections with CMOS. Furthermore, TSV arrays may impact the behavior of CMOS-based circuits by inducing thermo-mechanical stress in the front end of line~(FEOL) layer~\cite{tsv_design1}.
 %One the main issues with incorporating TSV with the CMOS integrated circuits is a lack of commercial EDA tool support for such hybrid designs. In some scenarios the the TSV connections might impact the behavior of the CMOS circuits by inducing thermo-mechanical stress in the front end of line~(FEOL) layer. The cross-talk and circuit mismatch are the other example of concerns in TSV-equipped CMOS designs. 
 
Three-dimensional TSV integration technology can enable a higher level of parallelism in highly-interconnected neuromorphic computing networks, with layered structures~\cite{tsv_neuro}. In such configuration, each layer contains one or multiple neural modules and TSV arrays represent inter-module connections. Given the huge number of inter-neuron connections, it is impractical to utilizes one TSV link per each neural connection, mainly because of TSV-to-TSV coupling~\cite{tsv_coupling}, crosstalk~\cite{tsv_design}, and density issues~\cite{tsv_neuro}. TSV technology offers opportunities for energy/performance improvements in large-scale neuromorphic design by reducing the network distance per spike~(parameter $d$ in the analysis presented in section.~\ref{sec:communication}) by having a higher radix router. Energy cost of communication networks is directly a function of the number of hops~(routers and interconnects) that neural spikes traverse from source to the destinations. Therefore, TSV-based neuromorphic systems are expected to offer better  energy cost compared to 2D structures. It is noted that the router's complexity in TSV-equipped hypercube neuromorphic designs increases due to extra links required per inter-layer router. In order to investigate the feasibility and impact of TSV technology on large-scale neuromorphic designs, it is paramount to have access to more detailed TSV models and design tools. 

%  As the neuromorphic systems are interconnect dominated design, the current TSV footprint can be more restricting in utilizing such technology for large-scale networks. The TSV size is currently in order of XXX. Despite the great potential in adapting TSV-based technology in neuromorphic designs, the requirements of such integration are not well investigated. Depending on how this technology is utilized, the requirements can be different. Given the relatively large TSV footprint, it is evident that it is impractical to use one TSV connection per each synaptic interconnection. Therefore, a multiplexing scheme is required to be combined with these type of connections. One possible way is to implement TSV connections per a group~(core) of neurons.  
\begin{figure}
\centering
\includegraphics[scale=0.52]{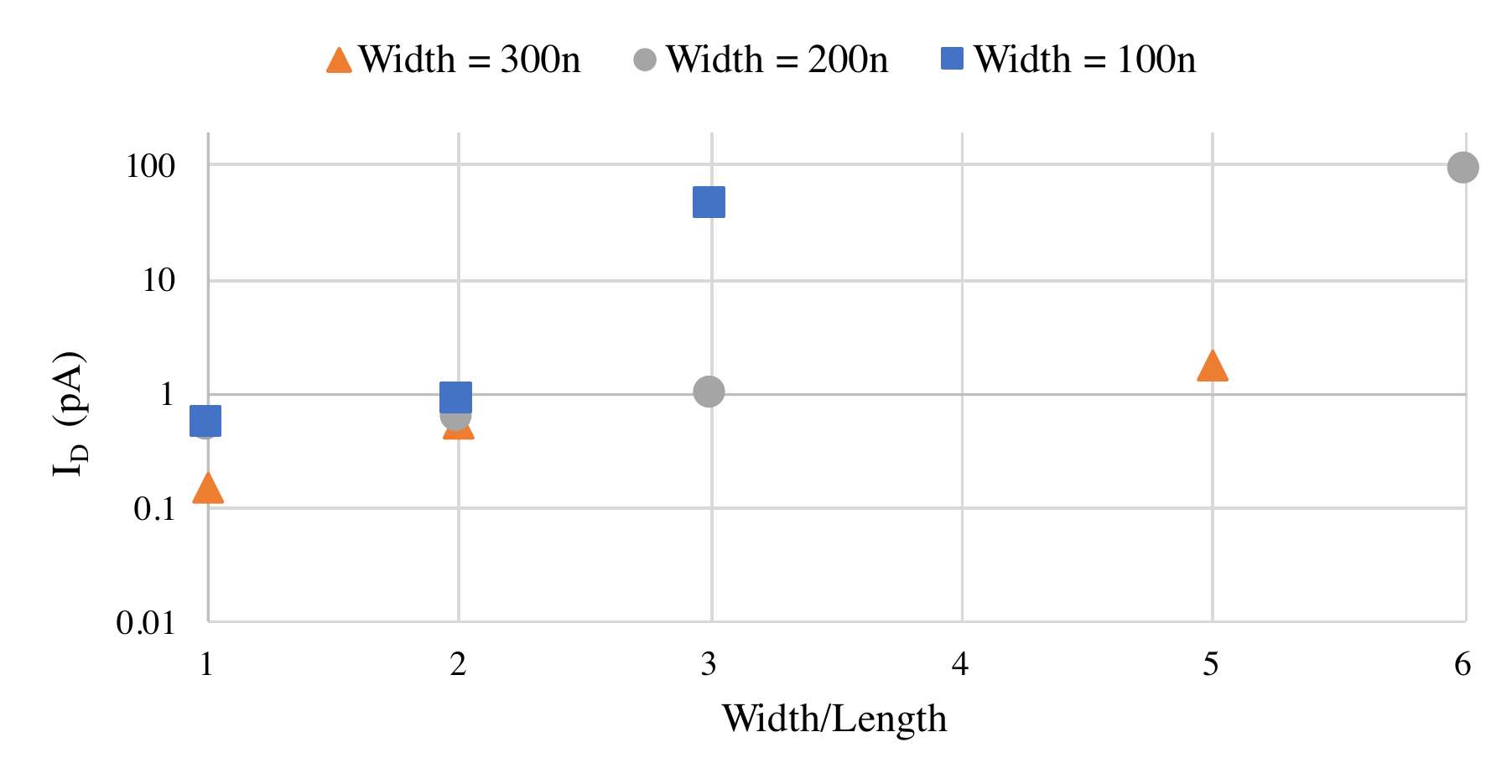}
\caption{Drain current~($I_{D}$) NMOS transistors with different sizes in 28\,nm process while $V_{gs}=0$, $V_{DS}=0.5V$.}
\label{fig:offcurrent}
\end{figure}

%\begin{figure}
%\centering
%\begin{subfigure}{.5\textwidth}
%  \centering
%\includegraphics[width=1.7in]{figs/nemsrelay.pdf}
%  \caption{NEM relay.}
%  \label{fig:nemsrelay}
%\end{subfigure}%
%\begin{subfigure}{.5\textwidth}
%  \centering
%\includegraphics[width=2.6in]{figs/nemsfets.pdf}
%  \caption{NEMS relays used as switches}
%  \label{fig:nems_config}
%\end{subfigure}
%\caption{(a): NEM relay showing a normally-open topology. (b) NEMS relays used %as switches~\cite{async_nems}. (a), (b) ``NMOS'' with
%  normally-open relay and normally closed relay; (c), (d)
%  ``PMOS'' with normally-open relay and normally-closed relay.}
%\label{fig:nemsfig}
%\end{figure}
\subsection{Nano-Electro-Mechanical Systems~(NEMS)}
\label{sec:nems}
Nano-electro-mechanical systems are miniaturized mechanical systems that use the mechanical properties of silicon (and other materials) in addition to the electronic properties to build circuits. NEMS has also been proposed as a way to reduce leakage currents in CMOS, and hence could benefit neuromorphic systems where leakage power dominates.

Excessive leakage current of field-effect transistors is a major obstacle in design of neuromorphic circuits in ultra deep sub-threshold CMOS processes~\cite{nems_neuro, cmos_leakage_neuro}. 
Device's off-current consume power even when the transistors are inactive. The problem aggravates in advanced CMOS processes~\textit{e.g. 28\,nm} where the off-current is in-negligible in comparison to on-current of the transistors, leading to an increase in total power dissipation. The power breakdown presented in Table~\ref{table:np_comparisons} shows that the static power consumption becomes a dominant factor of total power dissipation in large-scale custom neuromorphic systems, especially once computation power is reduced using emerging devices. For instance, the static power in TrueNorth~\cite{Merolla_etal14a}, a fully-digital design, accounts for 60\% of total power, at 20\,HZ firing rate. The power leakage issue in deep-submicron mixed-signal neuromorphic designs would be even more critical than digital designs~\cite{cmos_leakage_neuro}. Mixed-signal design typically occupy larger silicon area~(higher static power), due to their highly parallel structure. In addition, the leakage current may impact the behavior of analog sub-threshold circuit blocks. The neuromorphic sub-threshold CMOS neuron and synapse circuits typically operate within pico-ampere currents, while off-current of NMOS transistors in 28\,nm can be as high as pico amperes~(Fig.~\ref{fig:offcurrent}), even for long transistors~(e.g. Length=300\,nm), are integrated with digital switches. In other words, larger leakage current not only leads to higher power dissipations, but also narrows down the range of currents in which the neuromorphic sub-threshold circuits can reliably operate.

Nano electro-mechanical (NEM) switches have been recently been investigated as a potential alternative to CMOS~\cite{nems_neuro, nems_scaling, nems_chen, nems_scaling_2, nems_scaling_3}. %While NEMS are much slower than CMOS, their near-zero off current means that they enable significantly higher degrees of parallelism. 
The potential of NEM devices have been investigated for SRAM cells~\cite{nems_sram, nems_scaling,nems_sram1}, FPGAs~\cite{nems_fpga,nems_fpga1}, 3D integration~\cite{nems_thermal_3D}, asynchronous logic~\cite{async_nems}, etc. Hybrid CMOS-NEMS neuromorphic circuits, presented in~\cite{nems_neuro}, implement the leaky-integrate-fire~(LIF) neuron model~\cite{silicon_neuron} and take a step toward reducing the leakage power in a neuromorphic system. 
%Fig.~\ref{fig:nemsrelay} shows a normally-open NEM relay structure, which is a four-terminal switch consisting of a
%gate (G), drain (D), source (S), and body (B) terminals. When there is
%a large enough voltage difference (called the {\it pull-in voltage\/})
%across the gate and body ($V_{gb}$), the electrostatic force generated is large
%enough to cause the suspended structure to be pulled toward the body
%thereby connecting the source and drain terminals. When $V_{gb}$ is
%smaller than the pull-in voltage, the gate pulls away from the body
%due to the spring's restoring force. A similar principle is used for
%a normally closed NEM relay. 
Normally-open and normally-closed NEM relays can be used to build
replacements for NMOS and PMOS switches.
%Fig.~\ref{fig:nems_config}
%shows four configurations that implement the two types of transistors
%with the two different relay types.
A challenge with the ideal
scenario depicted here is that manufacturing uncertainty can require
{\it different\/} body voltages for different NEMS devices. Compared to digital CMOS switches, 
the switching delay of NEMS is high, and the number of time a device switches is limited before it fails. However, since neuromorphic systems operating on Biological time scales have extremely low activity, the reliability of NEM switches may not be as important a factor in this application domain compared to high-speed digital logic.

%Recent advances in semiconductor research offers integration of NEMS with CMOS. These devices have near-zero idle power consumption~\cite{nems_scaling}, thereby potentially enabling higher degrees of parallelism in a neuromorphic computing systems. The main drawbacks of NEM devices, namely limited switching lifetime and slow switching cycle, do not cause major issues in neuromorphic computing systems with biologically plausible time-constants. 
In addition to the neural computation circuits, the NEMS switches can be utilized in local communication blocks in order to alleviate the leakage power issues. Most custom-designed neuromorphic CMOS designs, discussed in Sec.~\ref{sec:cmos}, have utilized SRAM-based crossbar arrays to implement local connectivity. The SRAM blocks are typically designed using conventional 6-transistor circuits and suffer from high-leakage power consumptions. Within this context, future research works may focus on investigating the potential of NEMS switches in combination with CMOS, for optimally implementing local synaptic connections, as an alternative to high-leakage SRAM cells in large-scale neuromorphic designs.

\section{Discussion and Conclusion}
In recent years, research interests in exploring the potential of emergent nanoscale device technologies have been growing, as researchers have investigated their application to design more efficient memory and interconnections technologies. Large-scale neural networks are one of many applications that can significantly benefit from reliable emergent memory technologies. A large number of research articles have explored the potential of utilizing new memory devices e.g. memristors, to replace CMOS-based memories in implementing neuromorphic computing units e.g. synapses. The yield, scalability~(e.g. size and control overhead), and reliability issues are major obstacles in utilizing the emergent nanoscale devices with current technologies. Although large-scale neuromorphic hardware requires a considerable amount of storage in total, the storage is not centralized; rather, it consists of a large number of small blocks~(e.g. \cite{Merolla_etal14a}) that is distributed with the clusters of neurons and synapses. Hence, the control overhead of accessing a small memory block is non-trivial, so emerging memory technologies do not provide a means to tackle this issue unless they can also reduce this overhead. In order to efficiently utilize the emergent memory technologies in scalable neuromorphic systems, compact and power-efficient \emph{small\/} memories are needed.

Successful utilization of the emergent memories in large-scale neuromorphic designs is expected to further reduces power requirements for performing computation. Despite such potential improvement, the impact on total power consumption is debatable. In the recent neuromorphic developments, the power required for computation is not the dominant factor of the total power figure. Static power and communication power are a major fraction of total system power  in large-scale designs. Although the emergent memory devices partially help to reduce the static power by replacing high-leakage SRAM blocks, the communication power requirements would remain a major issue.

Emergent 3-D integration technologies such as TSV offer promising solutions to overcome the limitations facing current interconnection technology. TSV-based designs have been explored for variety of systems such as image sensors, memories, FPGAs and SoCs. The 3-D integration technologies have significant potential for highly-interconnected large neuromorphic networks by stacking multiple dies and reducing wire-complexity. Currently this technology is not easily accessible to designers, but this can change with time.

Because of their near-zero leakage current, CMOS-compatible NEMS devices could  alleviate the static power issues in CMOS circuit implementations. Hybrid CMOS-NEMS designs have been investigated for memories, FPGAs and 3D SoCs. These switches can be potentially be utilized in communication fabrics in neuromorphic networks. 

In this paper, we elaborated on the communication requirements of large-scale neuromorphic designs in terms of power and bandwidth requirements, provided a detailed comparison with conventional network-on-chip architecture, and discussed the potential improvements and issues in integrating the emergent nanoscale technologies in large-scale hardware neural networks. Research advances in emergent memory technologies can offer a route toward improved biological-scale neural computing systems only if they address communication and idle power requirements in addition to the computational requirements of such systems.

%\section*{References}

%\end{sortbibliography}

\end{document}